# Fabry-Pérot Interference, g-factor Anisotropy, and Gate-Tunable Quantum dot in Chiral Tellurium Nanowires

Suresh Ghimire[1], Mohammad Hafijur Rahaman[1], Nathan Tanner Sawyers[1], Madan Mohan Bhandari[1], Gokul Acharya[1], Syed Zulfiqar Hussain Shah[2,3], Iris Nandhakumar[3], Pawan Kumar[2], Zainul Aabdin[2], Hugh O. H. Churchill[1,4,5], and Dharmraj Kotekar-Patil[1,4,5*]

[1]*Department of Physics, University of Arkansas, Fayetteville, Arkansas 72701, United States*

[2]*Institute of Materials Research and Engineering, Agency for Science Technology and Research (A*STAR), Singapore 138634, Republic of Singapore*

[3] *Department of Chemistry, University of Southampton, Southampton SO17 1BJ, UK*

[4]*MonArk NSF Quantum Foundry, University of Arkansas, Fayetteville, Arkansas 72701, United States*

[5]*Arkansas Materials Institute, Fayetteville, Arkansas 72701, United States*

* *Corresponding author: dk030@uark.edu*

## ABSTRACT

Chiral materials with strong spin–orbit coupling offer a unique platform for exploring the interplay between topology, chirality, and quantum transport, yet the quantum coherent regime in elemental tellurium nanostructures remains largely unexplored. Here we demonstrate phase-coherent quasi-ballistic transport, anisotropic Zeeman spectroscopy, and gate-tunable quantum-dot formation in hydrothermally grown t-tellurium nanowires. Single nanowire field-effect transistors exhibit p-type transport with hole mobilities rising from ≈80 cm² V⁻¹ s⁻¹ at 210 K to ≈190 cm² V⁻¹ s⁻¹ at 1 K, consistent with a crossover from phonon-limited to Coulomb scattering dominated regimes near 50 K. Notably, devices segregate into two distinct regimes based on their room temperature two-terminal resistance: low-resistance devices (≤ 30 kΩ) exhibit Fabry–Pérot interference, whereas high-resistance devices (≥ 30 kΩ) display Coulomb-blockade behavior revealing a two-terminal resistance-driven transition between quasi-ballistic and strongly localized transport regimes. Zeeman spectroscopy in in-plane and out-of-plane magnetic fields yields highly anisotropic Landé g-factors (an in-plane $g_{\parallel} = 1.18$ and an out-of-plane $g_{\perp} = 18.41$) and directly resolves a spin-orbit energy gap $\Delta_{SO}$ = 0.864 meV from an avoided crossing. These results establish chiral tellurium nanowires as a versatile platform for gate-defined spin qubits exploiting large, tunable g-factors and for hybrid tellurium-superconductor architectures targeting Majorana zero modes in an elemental vdW system.

## Table of Contents (TOC) Graphic

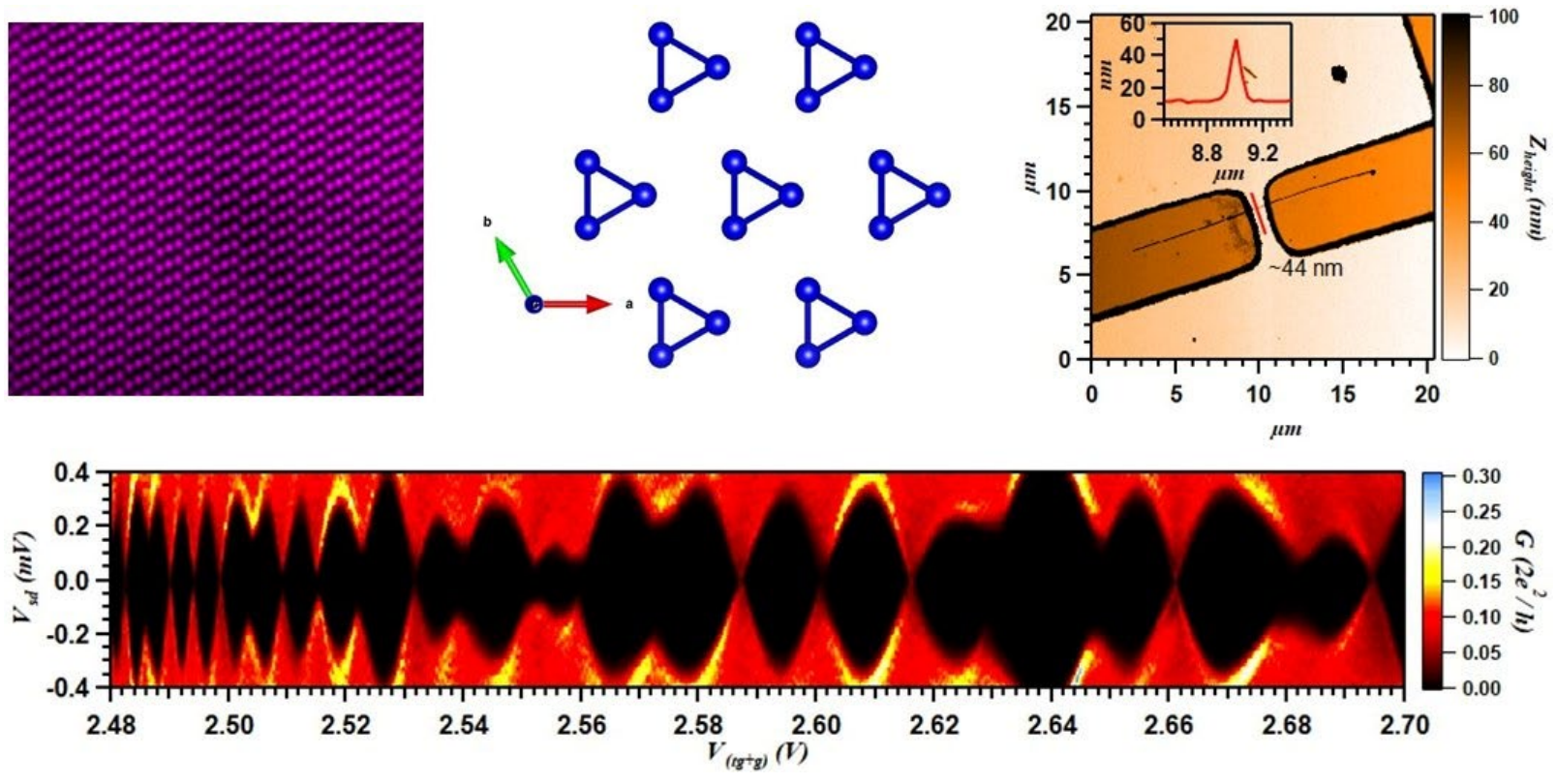

## INTRODUCTION

Trigonal tellurium (t-Te) is an elemental, chiral, narrow-bandgap (≈0.33 eV) p-type semiconductor whose lattice consists of helical atomic chains running along the c-axis and held together by van der Waals (vdW) interactions.[1,2] Its broken inversion symmetry combined with strong spin–orbit coupling (SOC) places Weyl nodes near the band edges and gives rise to a radial spin texture, non-reciprocal magnetoresistance, chirality-dependent transport, and a quantum Hall response of Weyl fermions in n-type tellurene.[3–7] Because the helical chains can be isolated down to a single-atom chain inside boron-nitride nanotube hosts, Te is a uniquely tunable platform for studying chirality and topology in reduced dimensions and a p-type channel material for sub-5-nm gate-all-around (GAA) transistors.[8–10] On the device side, Te has attracted intense interest as a high-mobility, air-stable, low temperature processable p-type semiconductor. Field-effect transistors from hydrothermally grown,[11] directly grown,[12] and PVD deposited[13,14] Te nanowires (NW) and flakes have demonstrated room-temperature hole mobilities ranging from ≈160 $cm^2 V^{-1} s^{-1}$ to ≈1485 $cm^2 V^{-1} s^{-1}$ (and up to ≈3500 $cm^2 V^{-1} s^{-1}$ at 2 K), with dual-gated junctionless Te NW FETs reaching ≈1390 $cm^2 V^{-1} s^{-1}$ at low temperature.[15] Across these reports, the temperature dependence of mobility typically shows a phonon-limited regime ($\mu \propto T^{-\alpha}$, $\alpha \approx 1$–1.5) above ~150 K crossing over to a low temperature Coulomb scattering regime below ~50 K, as expected for a clean vdW channel with low interface trap density.[12,15]

Despite this remarkable progress in classical transistor operation, the low-temperature quantum coherent regime in Te NWs remains relatievely unexplored. Phase-coherent transport signatures such as Fabry –Pérot (F-P) interference, universal conductance fluctuations (UCFs), and Aharonov–Bohm oscillations are routine benchmarks for (quasi-) ballistic semiconductor NWs such as InAs, InSb, and Ge/Si,[16–19] and have been observed in related telluride NWs (PbTe, SnTe, GeTe).[20–22] For elemental Te, only recently has phase-coherent transport been demonstrated. Rahaman et al. observed a crossover from Coulomb-blockade behavior to F-P interference in thin (13–60 nm) two-dimensional Te flakes at deep cryogenic temperatures with F-P visibility enhanced in the thinnest flakes.[23] To date, however, no such phase-coherent interference signatures has been reported in a Te NW geometry, where the quasi-one-dimensional confinement and helical chain alignment may dramatically modify mode-resolved transmission and spin texture. Similarly, Coulomb blockade effects in Te nanostructures have been relatively less explored. The Landé g-factor is a key parameter that encodes the spin–orbit dressed band structure, yet experimental values for nanoscale Te remain unmeasured. By contrast, related PbTe NW quantum dots (QDs) have yielded very large, anisotropic g-factors (≈20–44),[26] emphasizing how sensitive Zeeman spectra are to the host band structure and highlighting the value of analogous measurements in the elemental Te limit.

In this article we fabricate backgated and dual-gated devices from hydrothermally grown high-crystallinity Te NWs and perform transport measurements from ≈210 K down to base temperature (10 mK) in magnetic fields up to 4T. We (i) characterize the chiral helical-chain lattice by Raman spectroscopy and atomic-resolution STEM, confirming the t-Te crystal structure; (ii) extract the temperature dependence of the hole mobility from ≈80 $cm^2 V^{-1} s^{-1}$ at 210 K to ≈190 $cm^2 V^{-1} s^{-1}$ at 1 K and identify the operative scattering mechanisms (phonon-limited above ~50 K, Coulomb-limited below); (iii) demonstrate Fabry–Pérot interference as a direct fingerprint of phase-coherent quasi-ballistic transport in our Te NW; (iv) perform Zeeman spectroscopy in parallel and perpendicular magnetic fields to extract the in-plane and out-of-plane Landé g-factors and directly measure the spin-orbit gap ($\Delta_{SO}$) from an avoided crossing; (v) realize a clean, regularly periodic Coulomb blockade diamonds in single QD regime in a Te NW; and (vi) demonstrate electrostatic tunability of the QD size in a dual-gated geometry. Together, these results establish hydrothermal Te nanowires as a versatile platform for chirality- and spin – orbit driven quantum transport. Their large, anisotropic g – factors and strong spin – orbit coupling make them promising candidates for spin qubit applications. In addition, hybrid superconductor – Te nanowire devices provide a potential route toward realizing topological superconductivity in elemental tellurium systems.

## RESULTS AND DISCUSSION

### *Synthesis and structural characterization*

Te NWs were synthesized via a hydrothermal route adapted from the protocol of Zhang et al.[27] In a typical preparation, 20 mL of ethylene glycol was loaded into a three-neck round bottom flask connected to a standard Schlenk line. Polyvinylpyrrolidone (PVP, $M_w \approx 40{,}000$; 0.2 g), sodium hydroxide (NaOH; 0.6 g), and high-purity tellurium dioxide powder ($TeO_2$, 99.999%; 3 mmol) were sequentially introduced into the solvent under continuous stirring. The reaction mixture was then heated to 160 °C under an inert nitrogen atmosphere to remove dissolved oxygen and prevent oxidation of the tellurium precursors. Upon reaching the target temperature, 0.6 mL of hydrazine hydrate ($N_2H_4 \cdot H_2O$) was injected as a reducing agent to initiate the reduction of $Te^{4+}$ to $Te^0$. The reaction was maintained at 160 °C for 1 h, during which uniform, single-crystalline t-Te NWs nucleated and grew along the [0001] direction. After cooling naturally to room temperature, the as-synthesized NWs were collected by centrifugation, washed several times with methanol and deionized water to remove residual PVP surfactant and reaction byproducts, and redispersed in isopropanol for subsequent device fabrication.

The hydrothermal route consistently yields NWs with typical lengths of ~15 µm and diameters ranging up to 100 nm, as shown in the low-magnification TEM image of Figure 1a. The NWs exhibit a uniform, straight morphology with minimal branching or aggregation, indicating that the PVP surfactant effectively controls the anisotropic growth along the c-axis. The crystal structure of t-Te is illustrated in Figure 1b,c. Viewed along the [0001] axis (Figure 1b), the lattice consists of a triangular sub-lattice of vdW-bonded helical chains arranged in a hexagonal pattern. The side view (Figure 1c) shows three parallel, three-atom period helices propagating along the c-axis, held together by weak vdW interactions which is the structural origin of the chirality and anisotropic transport properties characteristic of t-Te. The Raman spectrum of a single Te NW (Figure 1d) shows three characteristic vibrational modes: the $E_1$ (transverse, bond-bending) mode at ~93 $cm^{-1}$, the $A_1$ (symmetric intrachain chain-breathing) mode at ~122 $cm^{-1}$, and the $E_2$ (asymmetric stretching along c) mode at ~141 $cm^{-1}$. These mode positions and assignments are in excellent agreement with the established Raman signature of bulk t-Te (92, 120, 141 $cm^{-1}$)[28–30] confirming both the high crystalline quality and the t-Te phase of the as-grown NWs. Atomic-resolution HAADF-STEM imaging (Figure 1e,f) directly resolves the periodic atomic registry of the helical chains in the NW confirming the [0001] growth direction inferred from the low-magnification TEM and ruling out significant stacking faults or amorphous content. The combination of long, uniform morphology and high crystallinity make these hydrothermally grown Te NWs an excellent platform for low-temperature quantum transport experiments.

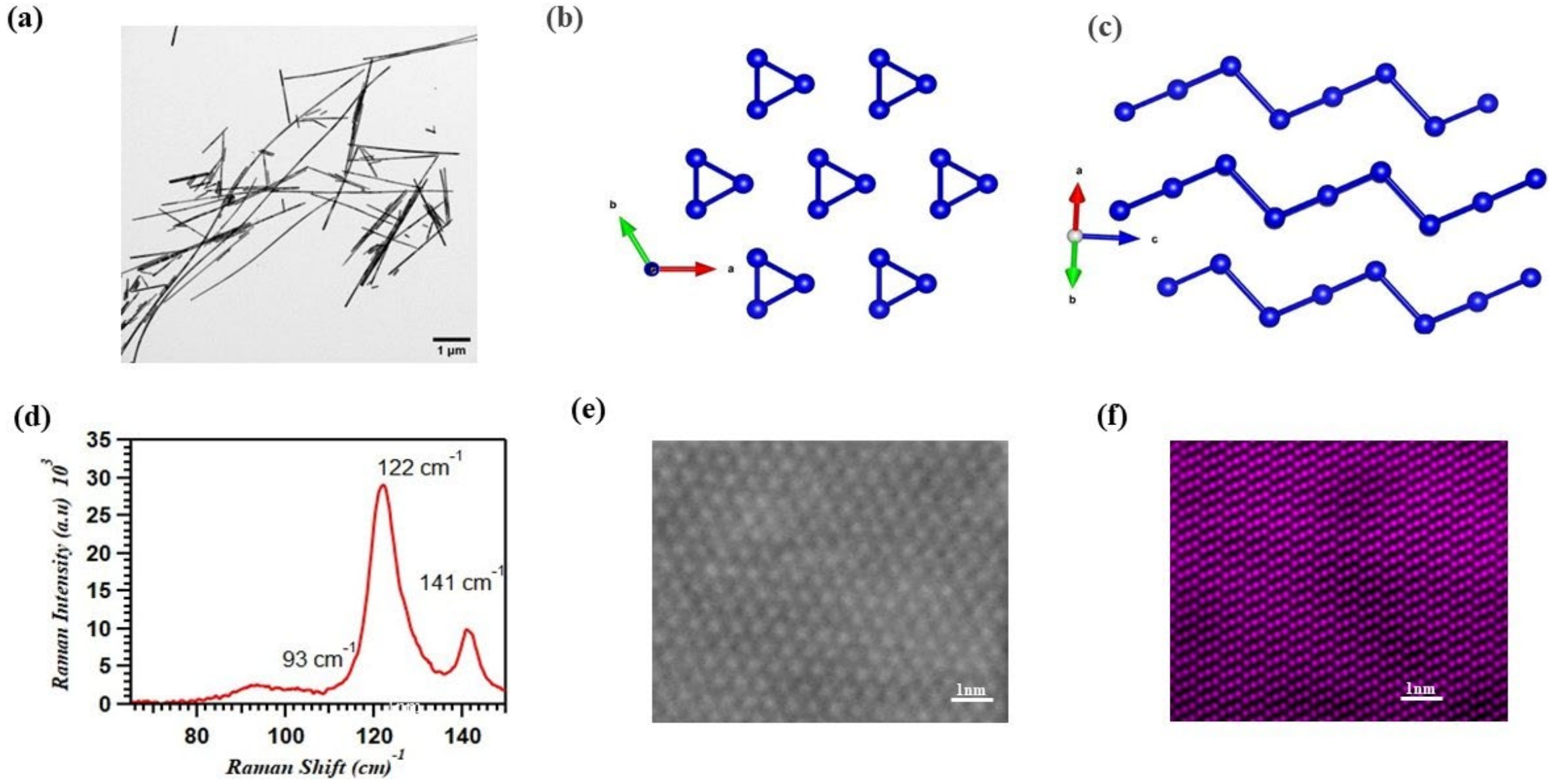

**Figure 1. Hydrothermal Te NWs: morphology, crystal structure and Raman signatures.** (a) Low-magnification TEM image of as-synthesized t-Te NWs. (b) Top view of the t-Te crystal along the [0001] axis showing the triangular arrangement of vdW-bonded Te helices. (c) Side view showing three parallel Te helical chains propagating along c. (d) Raman spectrum of a single Te NW with the characteristic $E_1$ (≈93 cm$^{-1}$), $A_1$ (≈122 cm$^{-1}$) and $E_2$ (≈141 cm$^{-1}$) modes of t-Te. (e) Atomic-resolution HAADF-STEM image of a Te NW resolving the helical-chain atomic arrangement. (f) False-color rendering emphasizing the periodic t-Te sub-lattice.

### *Field-effect transport and temperature dependence of the hole mobility*

To characterize the electrical transport properties of the hydrothermally grown Te NWs, we performed two-terminal conductance measurements on individual NWs contacted with source and drain electrodes patterned by photolithography. The degenerately doped Si substrate underneath the 285 nm $SiO_2$ dielectric served as a global backgate ($V_g$). Transfer characteristics $G(V_g)$ were recorded over a wide temperature range from $T = 210$ K down to $T = 1$ K, as shown in Figure 2a. Across the entire temperature range, the devices exhibit clear p-type field-effect behavior consistent with hole transport in the t-Te valence band.[11,12,15] In our Te nanowire devices, a clear metal–insulator transition (MIT) is observed (Figure 2a), where the current decreases with increasing temperature at negative gate voltages (0V and below metallic behavior) and shows conventional semiconducting temperature dependence at more positive gate voltages. This MIT can be understood within a classical percolation framework, where insufficient screening at low carrier density leads to an inhomogeneous potential landscape and the formation of insulating puddles. At higher carrier densities, enhanced screening restores metallic transport.

To quantify the carrier mobility, we extract the field-effect hole mobility $\mu$ from the linear regime of the transconductance using the standard expression

$$\mu = (L^2 / C_g) \times (dG / dV_g) = (g_m \times L^2) / (C_g),$$

where $g_m = dG/dV_g$ is the transconductance evaluated in the linear regime of the $G(V_g)$ trace, $L$ is the channel length, and $C_g$ is the gate capacitance of the NW. The extracted $\mu$ ($T$) for the device shown in Figure 2a is plotted in Figure 2b. The mobility rises monotonically from $\approx$ 80 $cm^2$ $V^{-1}$ $s^{-1}$ at 210 K to a peak value of ≈190 $cm^2$ $V^{-1}$ $s^{-1}$ at 1 K, exhibiting a power-law temperature dependence $\mu \propto T^{-\alpha}$ over an extended temperature range.

Fitting the high-temperature regime ($T > 50$ K) to the power-law expression yields an exponent $\alpha = 0.6$, shown as the solid line in Figure 2b. This phonon-limited behavior with $\mu$ decreasing as temperature rises is characteristic of a semiconductor channel where carrier transport is governed by electron – phonon scattering rather than by static disorder. The extracted exponent $\alpha = 0.6$ is notably smaller than the canonical value of $\alpha = 1.5$ expected for pure acoustic-deformation potential scattering in bulk 3D semiconductors,[15] and lower than the $\alpha \approx 1.47$ reported for dual-gated junctionless Te NWs.[15] Instead, the measured exponent is remarkably close to the value expected when carrier mobility is limited predominantly by optical phonon scattering. Theory predicts $\mu \propto T^{-0.5}$ for optical phonon limited transport, and a closely matching exponent of $\alpha = 0.6 \pm 0.1$ has previously been reported for InAs NW and attributed specifically to optical phonon scattering.[31] Below $T \approx 50$ K, $\mu$ deviates from the power-law fit and saturates near ≈190 $cm^2$ $V^-$ [1] $s^{-1}$ indicates that charged-impurity and surface scattering contribute only minimally in our Te NWs. Instead, the mobility plateau reflects a residual, temperature-independent scattering background establishing the high crystalline quality and low-disorder character of these NWs.

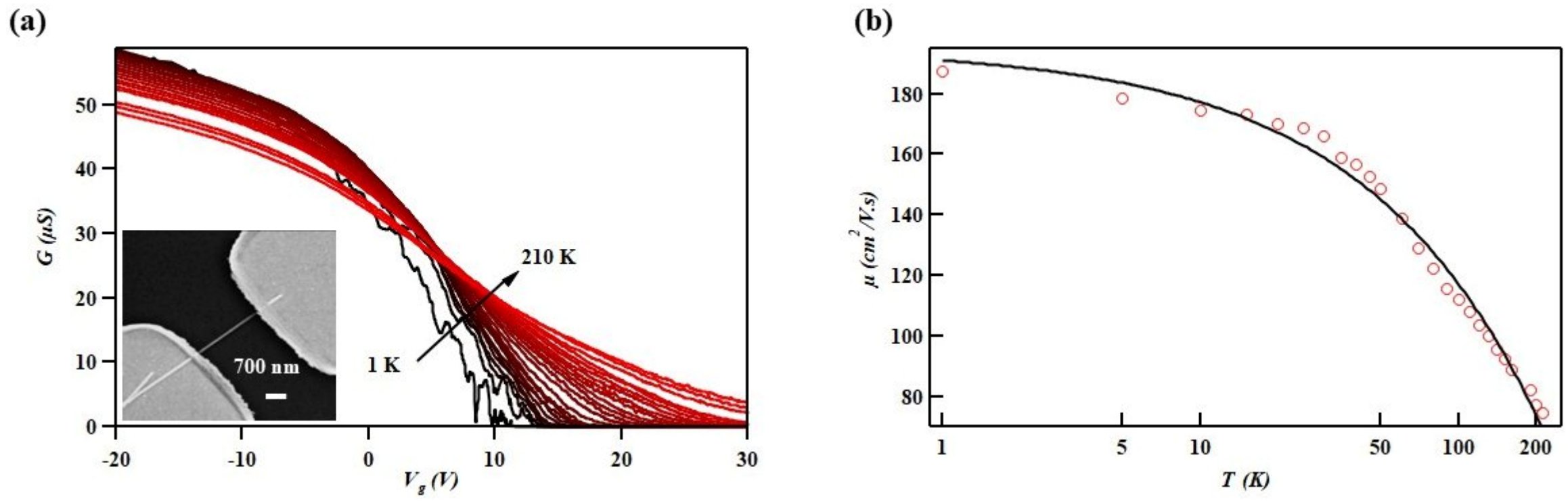

**Figure 2. Temperature-dependent transport of a single Te NW FET.** (a) Conductance $G$ versus back-gate voltage ($V_g$) at temperatures from 1 K (black) to 210 K (red). Inset: SEM image of the back-gated Te NW device. (b) Field-effect hole mobility ($\mu$) extracted from the linear regime versus temperature ($T$). The solid line is a fit consistent with phonon limited transport ($\mu \propto T^{-\alpha}$) above ≈50 K, crossing over to a mobility saturation at low-temperature indicating that the scattering from charged impurities is not a dominant factor in our device.

Across the ten devices characterized in this study, we observe a systematic segregation into two regime based on the room temperature two-terminal resistance $R_{RT}$. Devices with $R_{RT} \leq 30$ kΩ consistently exhibit Fabry – Pérot (FP) interference at millikelvin temperatures, whereas devices with $R_{RT} \geq 30$ kΩ display Coulomb blockade (CB) behavior. Notably, the observed resistance of ∼30 kΩ lies near the expected crossover between ballistic and localized transport regimes, which occurs on the order of the quantum resistance $h/e^2$ (~ 25.8 kΩ). Table 1 summarizes the measured resistance and observed transport regime for each device. This resistance driven transport behavior reveals that the low total resistance enables quasi-ballistic transmission (FP regime), while higher total resistance lead to strong localization and single-electron charging effects (CB regime).

**Table 1.** Room temperature two-terminal resistance ($R_{RT}$) and observed transport regime (Fabry–Pérot interference or Coulomb blockade) for ten Te NW devices. Devices with $R_{RT} \leq 30$ kΩ consistently show FP, while devices with $R_{RT} \geq 30$ kΩ show CB.

| Device | R(RT) (kΩ) | Observation |
|---|---|---|
| Device1 | 28 | FP |
| Device2 | 21 | FP |
| Device3 | 28 | FP |
| Device4 | 23 | FP |
| Device5 | 25 | FP |
| Device6 | 51 | CB |
| Device7 | 44 | CB |
| Device8 | 41 | CB |
| Device9 | 30 | CB |
| Device10 | 44 | CB |

***Fabry – Pérot interference and phase coherent transport.***

At millikelvin temperatures, devices with two-terminal resistance $R_{RT} \leq 30$ kΩ (Table 1, Devices 1–5) exhibit conductance oscillations as a function of the backgate voltage, signaling the onset of coherent quantum transport. Figure 3a shows a 2D colormap of the differential transconductance $dG/dV_g$ as a function of $V_g$ and source-drain bias $V_{sd}$. Zooming into a narrow gate window around $V_g \approx -4$ V (Figure 3b), the oscillations evolve into a characteristic conductance checkerboard pattern, a hallmark of Fabry–Pérot (FP) interference in a coherent one-dimensional cavity.[16–18,23,34] This checkerboard signature, consisting of alternating lobes of positive and negative $dG/dV_g$ arranged in a crisscross pattern, has been widely observed in ballistic carbon nanotubes,[34] InAs NWs,[17,18] Ge/Si core-shell NWs,[16] and, very recently, in two-dimensional Te flakes.[23]

The physical mechanism underlying the FP checkerboard is analogous to an optical Fabry – Pérot etalon, but here the partially reflecting mirrors are the scattering centers within the NW channel, most likely crystallographic defects or local potential fluctuations. Two such scatterers separated by a distance $L_c$ defines a one-dimensional cavity in which holes undergo elastic, phase-coherent multiple reflections. A hole traversing this cavity accumulates a round-trip phase $\varphi = 2k_F L_c$, where $k_F$ is the Fermi wavevector. When $\varphi = 2\pi n$ ($n$ integer), the multiple reflected partial waves interfere constructively and produce a conductance maximum; when $\varphi = (2n + 1)\pi$, destructive interference yields a conductance minimum. Sweeping $V_g$ tunes $k_F$ and thus the round-trip phase, producing the periodic oscillations. Sweeping $V_{sd}$ shifts the injection energy and therefore the wavevector, tilting the constructive/destructive conditions to generate the diagonal crisscross lines of the checkerboard. The diagnostic confirmation of FP interference is the π-phase reversal of $dG/dV_g$ oscillations at finite $V_{sd}$ relative to $V_{sd} = 0$ as shown in Figure 3c, line cuts at $V_{sd} = 0$ (green) and $V_{sd} = 0.1$ mV (red) are phase-shifted by π, unambiguously distinguishing FP resonances from Coulomb blockade phenomena (where no such reversal occurs).

From the checkerboard pattern, the cavity length $L_c$ and the coherent transport length scales can be extracted. The quantized energy level spacing of the FP cavity is given by

$$\Delta E = \hbar^2\pi^2 / (2mL_c^2),$$

where $m$ is the effective mass ($0.1m_0$) and $L_c$ is the cavity length. $\Delta E$ is read directly from the $V_{sd}$ extent of the checkerboard lobes, which in our data yields an energy scale $\Delta E \approx$ 50–120 μeV across the accessible gate voltage range. Applying the above relation, we extract a $L_c \approx$ 177–274 nm. Importantly, this length is shorter than the channel dimensions of the measured devices, which range from a few hundred nanometers to approximately one micrometer. This confirms that the FP cavity is not bound by the contacts but instead by internal elastic scatterers within the NW consistent with crystallographic defects or local potential inhomogeneities in the Te lattice. The cavity length is physically bound by two characteristic length scales. The elastic mean free path $\ell_e$ (lower bound) and the phase-coherence length $\ell_\varphi$ (upper bound), such that $\ell_e \leq L_c \leq \ell_\varphi$. The lower bound arises because the FP cavity requires at least one elastic scattering path of order $\ell_e$ between the two reflectors for the resonance conditions to be established. The upper bound is set by $\ell_\varphi$ because inelastic dephasing beyond this length destroys the phase coherence needed for constructive and destructive interference.

Since $L_c \approx$ 173–254 nm is shorter than the full device length, the overall device level transport involves multiple such quasi-ballistic FP segments in series, but within each segment the carriers maintain phase coherence and undergo resonant transmission. This quasi-ballistic character is consistent with the relatively low-disorder transport established in mobility saturation at low T and with the crystalline structure confirmed by STEM (Figure 1e,f).

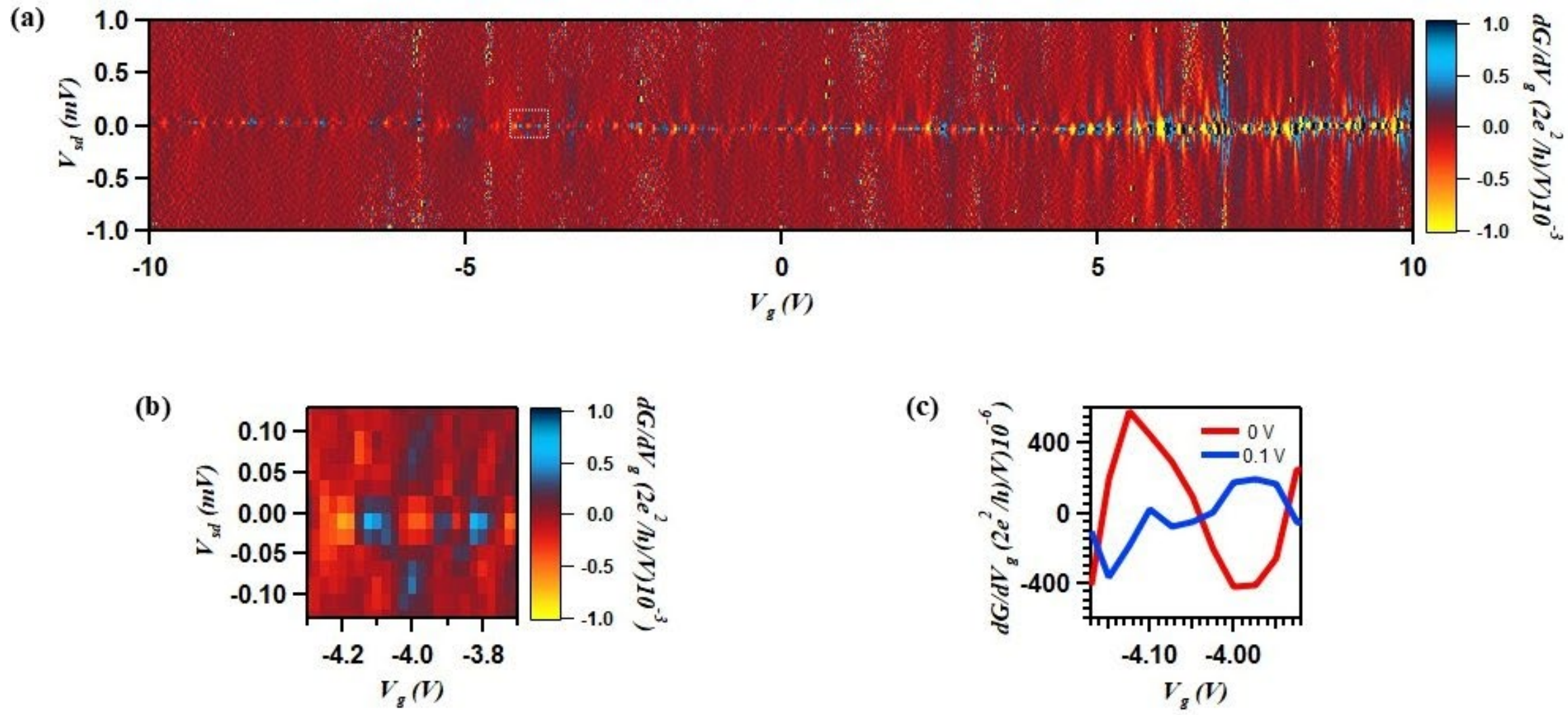

**Figure 3. Fabry–Pérot interference in a single TeNW.** (a) Transconductance as a function of source-drain bias and back-gate voltage measured at base temperature of 10mK. (b) Zoom-in around $V_g \approx -4$ V (dashed white box in (a)) revealing the transconductance checkerboard pattern with sign alternating $dG/dV_g$ lobes. (c) Line cuts of $dG/dV_g(V_g)$ at $V_{sd}$ = 0 V (green) and $V_{sd}$ = 0.1 mV (red). The phase reversal of the oscillations with $V_{sd}$ is the fingerprint of FP resonances that establishes phase coherent quasi-ballistic transport in the Te NW.

***Zeeman spectroscopy: g-factor anisotropy and spin-orbit gap.***

To probe the spin structure of the carrier states, we measured the conductance oscillations of the device as a function of both the backgate voltage $V_g$ and an applied magnetic field $B$, oriented either in-plane (along NW c-axis) or out-of-plane (perpendicular to both the NW axis and the substrate). As shown in Figure 4a, the conductance oscillation peaks evolve continuously with magnetic field and the conductance peaks around $V_g \sim 6.7$ V, undergo a splitting as $B$ increases. This splitting is the Zeeman effect i.e. magnetic field lifts the spin degeneracy, separating it into spin-up and spin-down branches with an energy difference $\Delta E_Z = g \mu_B B$, where $g$ is the effective Landé $g$-factor and $\mu_B$ is the Bohr magneton. By tracking the field dependent peak splitting in the in-plane configuration (Figure 4a) and fitting $\Delta E_Z$ versus $B$ to a linear dependence (Figure 4b), we extract an effective in-plane g-factor g = 1.18 ± 0.10. Because the in-plane field has an angle of ≈ 43° with the NW c-axis, this value represents an effective g-factor that mixes the c-axis and in-plane-transverse components of the g-tensor. In either case it remains small (≈1.2), well below the free-electron value of g ≈ 2.

For the out-of-plane field configuration (Figure 4c,d), the same analysis yields a larger perpendicular g-factor $g_\perp = 18.41 \pm 0.62$, more than an order of magnitude larger than $g_\parallel$. Notably, in the perpendicular field measurement we observe, in addition to the linear Zeeman splitting, a pronounced anti-crossing of conductance peaks (zoomed-in region in Figure 4c). As two spin-split levels are brought into near degeneracy by the field, they do not cross but instead repel each other, opening an avoided anticrossing. This level repulsion is a direct signature of spin-orbit coupling which mixes states of nominally opposite spin and forbids crossing. The magnitude of the anti-crossing gap $2\Delta_{SO} = 0.772$ meV (extracted from the minimum level separation in Figure 4c) provides a direct transport-based measure of the spin-orbit interaction strength in the Te NW.

The spin-orbit energy $\Delta_{SO}$ = 0.386 meV extracted here can be directly benchmarked against other one-dimensional semiconductor platforms that are actively pursued for spin-orbit physics and topological superconductivity. InSb NWs, which possess one of the strongest known spin-orbit interactions among III–V semiconductors, exhibit spin-orbit energies of ≈0.25–1 meV and effective g-factors up to ≈50.[19,25] InAs NW QDs show somewhat weaker SOC with spin-orbit energies typically ≈0.1–0.5 meV and gate-tunable g-factors in the range ≈2–15.[33] Hole systems such as Ge/Si core-shell NWs display strongly anisotropic, electrically tunable hole g-factors,[32] with spin-orbit energies on the order of ≈0.1–1 meV depending on confinement. PbTe NW QDs have yielded large, anisotropic g-factors of ≈20–44,[26] and carbon nanotubes exhibit much weaker SOC (≈0.01–0.1 meV). The presence of a strong spin-orbit coupling ($\Delta_{SO}$ ≈ 0.386 meV) with g-factor anisotropy is particularly interesting ingredients required for realizing topological superconductivity and Majorana zero modes in semiconductor NWs. Crucially, unlike in InSb or InAs, where strong SOC depends on the heavy-atom composition and must be enhanced by external electric fields, the spin-orbit coupling in Te NWs in addition to heavy-atom composition, is intrinsic to the chiral helical chain crystal structure offering a fundamentally richer material platform for spin-orbit driven quantum devices.

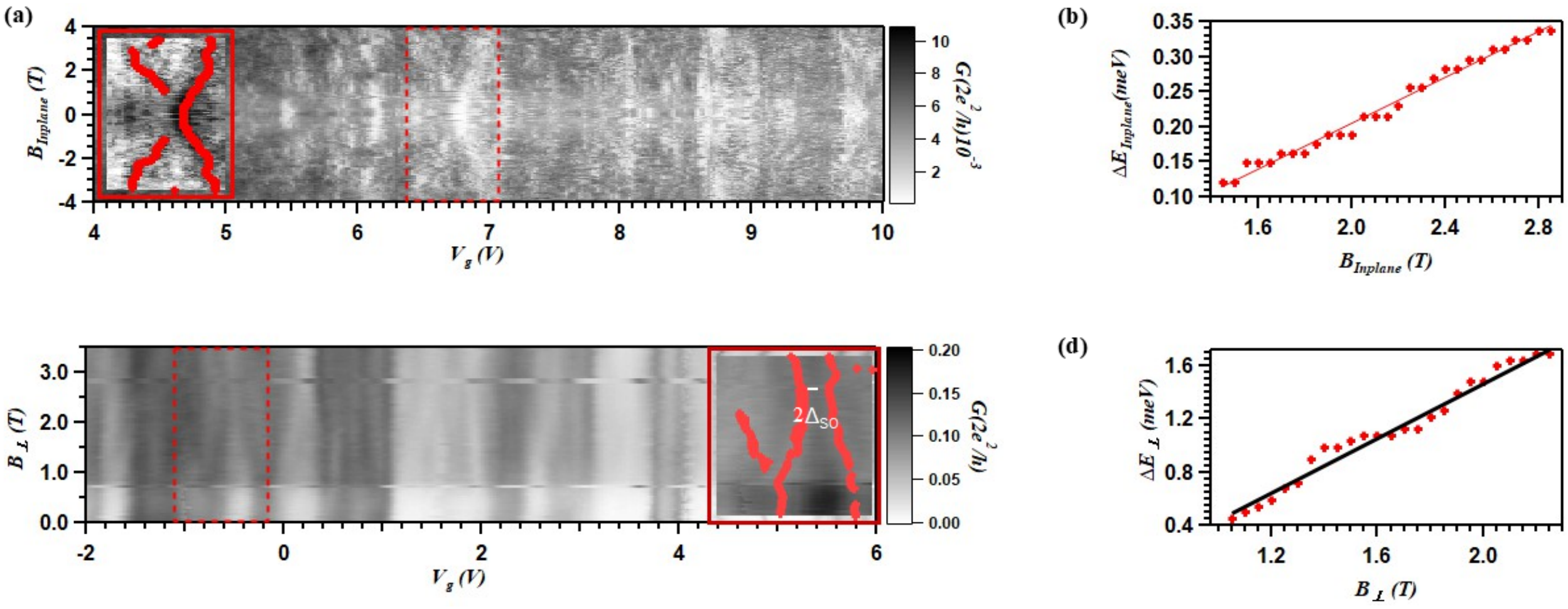

**Figure 4. Zeeman spectroscopy, g-factor anisotropy and spin-orbit gap in a TeNW.** (a) Conductance $G$ versus $V_g$ and in-plane magnetic field $B_{Inplane}$ (parallel to NW axis) up to ±4 T at base T. Red traces overlaid on the data track display a single Zeeman split conductance feature in the dashed box. (b) Extracted Zeeman splitting $\Delta E_{Inplane}$ versus $B_{Inplane}$, the linear fit (red) gives the effective in-plane Landé g-factor $g_{/\!/}$ = 1.18 ± 0.10 (B at ≈43° to the NW c-axis). (c) $G$ versus $V_g$ and perpendicular magnetic field $B_{\perp}$. The zoom (red box, right) shows an avoided crossing whose zero-field gap $2\Delta_{SO}$ = 0.772 meV directly measures the spin-orbit coupling strength. (d) Extracted Zeeman energy $\Delta E_{\perp}$ versus $B_{\perp}$. The linear fit yields the out-of-plane g-factor $g_{\perp}$ = 18.41 ± 0.62.

***Single quantum dot Coulomb blockade regime.***

In contrast to the FP devices, those with a two-terminal resistance $R_{RT}$ > 30 kΩ (Table 1 Devices 6 –10) exhibit a distinct phenomenon at low temperatures. As shown in Figure 5a, the conductance displays strong, highly periodic oscillations as a function of the backgate voltage $V_g$ with more than twelve equally spaced peaks of nearly uniform amplitude. Mapping the conductance as a function of both $V_g$ and source-drain bias $V_{sd}$ (Figure 5b) reveals that these conductance oscillations evolve into diamond-shaped domains of suppressed conductance, a hallmark of Coulomb blockade.[24,25] Inside each diamond the conductance is fully suppressed ($G \rightarrow 0$), reflecting the Coulomb gap that prevents the addition or removal of a single hole from the QD, whereas outside the diamonds sequential single-hole tunneling restores finite conductance. The observation of more than fifteen nearly identical Coulomb diamonds with uniform height (charging energy $E_C = e^2/C_\Sigma$, where $C_\Sigma$ is the total capacitance) and uniform gate voltage spacing ($\Delta V_g = e/C_g$) demonstrates the formation of a well-defined and a stable single QD within the Te-NW channel.

A notable feature of the data is the regularity of the diamond pattern i.e. successive diamonds show no measurable change in size or shape. In semiconductor QDs, when the single-particle level spacing $\delta E$ is comparable to the charging energy $E_C$, successive addition energies alternate between $E_C + \delta E$ (adding a carrier to a new orbital) and $E_C$ (filling a spin-degenerate orbital), producing a visible uneven modulation of the diamond size.[24,25] The absence of any such alternation in Figure 5b indicates that $\delta E \ll E_C$, placing this QD in the so-called "metallic" Coulomb blockade regime where the level spacing is negligible relative to the charging energy and every single hole addition costs the same electrostatic energy $E_C$ regardless of orbital index.

The gate voltage period $\Delta V_g$ of consecutive Coulomb peaks is directly related to the gate capacitance through $\Delta V_g = e/C_g$ where a larger period implies a smaller $C_g$ and vice-versa. In a cylindrical NW geometry, $C_g$ is linked to the quantum dot length $L_{dot}$ through the cylinder-on-plane capacitance model:[35]

$$L_{dot} = [ln(2h_{oxide} / r_{NW}) / (2\pi\varepsilon)] \times C_g,$$

where $h_{oxide}$ is the gate-dielectric thickness, $r_{NW}$ is the NW radius, and $\varepsilon = \varepsilon_0\varepsilon_r$ is the permittivity of the dielectric. Using $r_{NM}$ = 36 nm (AFM thickness measurement), $h_{oxide}$ = 285 nm, we get $L_{dot}$ as ~ 442.9 nm in good agreement of the device length (~ 500 nm). A rough estimate supports a picture that a single QD spans the entire NW segment between source and drain contacts.

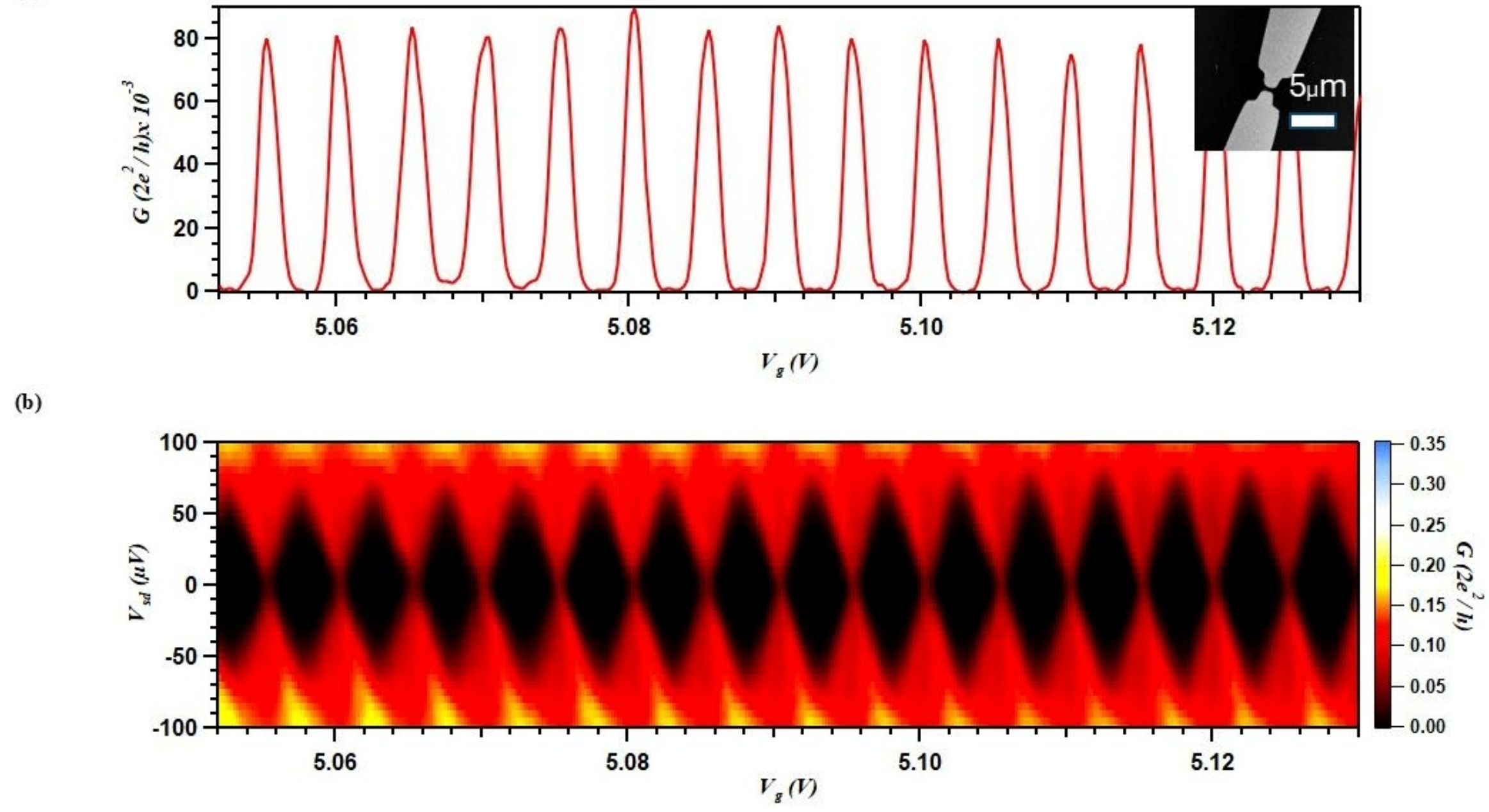

**Figure 5. Periodic Coulomb blockade single quantum dot regime in a Te NW.** (a) Periodic Coulomb-peak oscillations of $G$ versus $V_g$ in the window $V_g$ = 5.05–5.13 V, showing >15 regularly spaced peaks of approximately uniform height (peak G ≈ 80 × $10^{-3}$ × $2e^2/h$). Inset: SEM image of the device (scale bar 5 µm). (b) Bias spectroscopy $G(V_{sd}, V_{bg})$ over the same gate window reveals >15 nearly identical Coulomb diamonds. The uniform diamond geometry gives a single charging energy $E_C$ and confirms a well-defined single QD operating in the strong-coupling, low-disorder limit.

***Electrostatically tunable quantum dot in a dual-gated Te NW.***

To explore gate controlled tunability of the QD confinement, we fabricated a device with both a global backgate and a local top gate ($V_{tg}$) separated from the NW by a 40-nm thick hexagonal boron nitride (hBN) flake that serves as gate dielectric for the top gate. A scanning electron micrograph of a dual-gated device is shown in Figure 6a. Figure 6b displays the conductance $G$ as a function of $V_{sd}$ and the combined gate voltage $V_{(tg+g)}$, where both gates are swept simultaneously. The data reveal a continuous evolution of Coulomb diamonds across the accessible voltage window. At the left hand side of the map (lower combined gate voltage), the diamonds are closely spaced with a small gate voltage period $\Delta V_g$, whereas sweeping to the right (higher gate voltage) the period gradually increases and the diamonds widen. Simultaneously, the charging energy $E_C$ read from the diamond height in $V_{sd}$ increases from ≈0.3 meV at the left to above 0.4 meV at the right side of the map. Both trends are signatures of a QD whose size shrinks as the combined gate voltage is varied.

The smooth, monotonic increase of $\Delta V_g$ across the map therefore tracks a gradual reduction of $C_g$. Figure 6c plots the extracted dot radius (using the expression for $L_{dot}$ above) and gate capacitance $C_g$ as a function of the gate voltage spacing $\Delta V_g$. Both quantities decrease monotonically: $C_g$ drops from ≈50 aF to ≈10 aF, and the corresponding dot radius shrinks from ≈40 nm down to ≈10 nm. When both the top gate and back gate are swept simultaneously, their electric fields act from opposite sides of the NW pushing the hole density away from both gate dielectric interfaces. This dual sided electrostatic confinement effectively compresses the conducting channel into a progressively narrower core region of the NW, reducing the effective radius of the QD. As the dot radius shrinks, $C_g$ decreases and the total dot capacitance $C_\Sigma$ drops, driving the charging energy $E_C = e^2/C_\Sigma$ upward from ≈0.3 meV to above 0.4 meV.

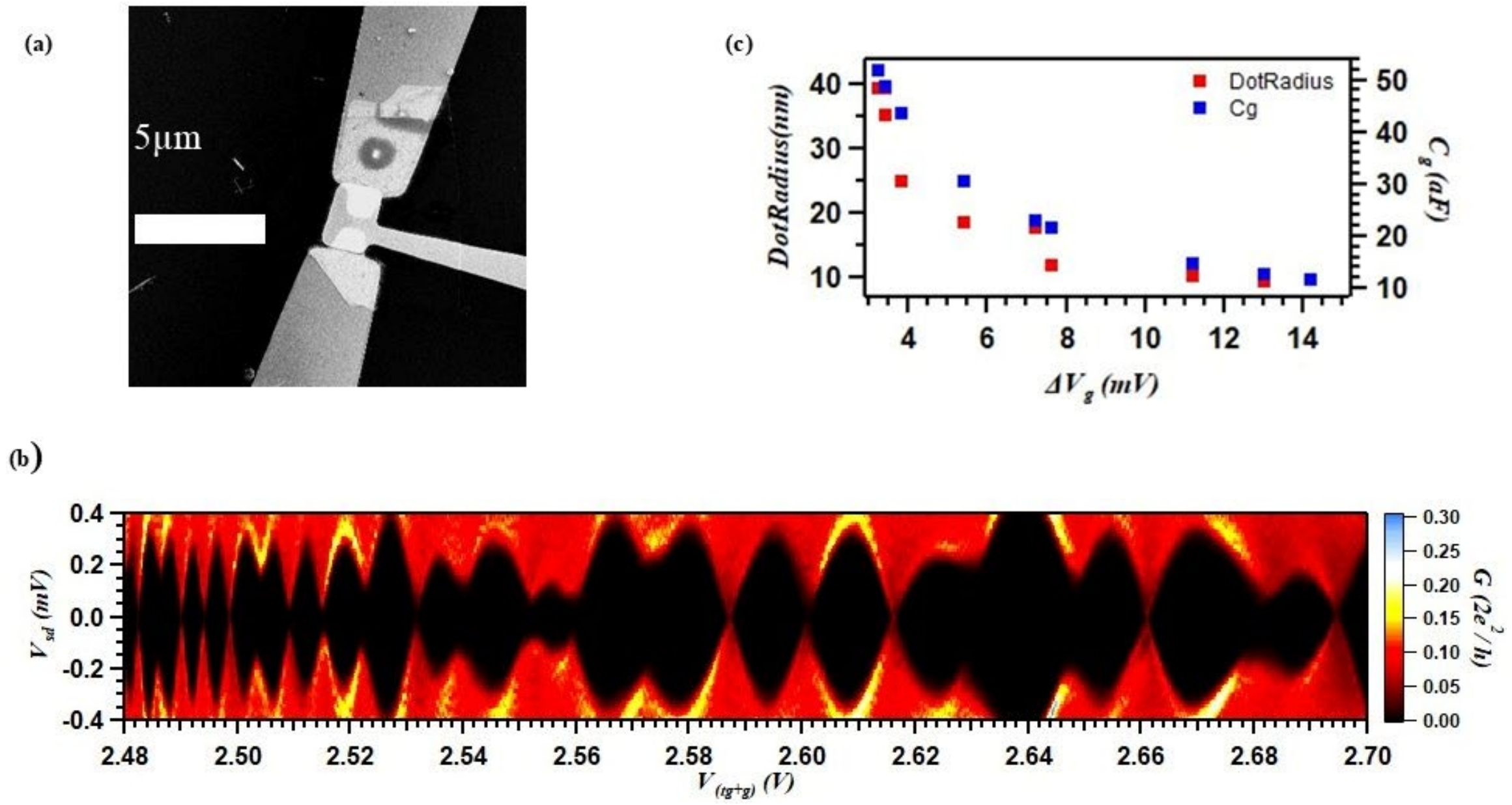

**Figure 6. Gate-controlled quantum-dot size in a dual-gated TeNW.** (a) SEM image of the dual-gated Te NW device with source, drain, top-gate, and backgate electrodes (scale bar 5 µm). (b) Coulomb diamond map $G(V_{sd}, V_{(tg+g)})$ over the differential gate voltage range $V_{(tg+g)}$ = 2.48–2.70 V, with strong modulation of Coulomb diamond periodicity. (c) Extracted dot radius (red squares) and gate capacitance $C_g$ (blue squares) versus the gate-voltage difference $\Delta V_g$. Both parameters decrease monotonically from ≈40 nm and ≈50 aF to ≈10 nm and ≈10 aF, demonstrating continuous electrostatic compression of the QD by tuning the differential confinement potential between the top and back gates.

## CONCLUSIONS

In summary, we have established hydrothermally grown t-tellurium NWs as a versatile platform for low-dimensional quantum transport and spin-orbit physics. Single NW FETs exhibit p-type hole mobilities rising from ≈80 $cm^2 V^{-1} s^{-1}$ at 210 K to ≈190 $cm^2 V^{-1} s^{-1}$ at 1 K with optical-phonon-dominated scattering ($\mu \propto T^{-0.6}$) and low-disorder saturation below 50 K, while a systematic study of ten devices reveals a contact resistance threshold at $R_{RT} \approx 30$ kΩ separating Fabry–Pérot interference (with cavity length $L_c \approx$ 173 –254 nm confirming quasi-ballistic transport) from Coulomb blockade. Zeeman spectroscopy yields strongly anisotropic g-factors ($g_{/\!/}$ = 1.18 ± 0.10 and $g_{\perp}$ = 18.41 ± 0.62) and a spin-orbit gap $\Delta_{SO}$ = 0.386 meV comparable to advanced material systems like InSb. In the Coulomb-blockade regime we realize clean metallic QD and in a dual-gated geometry we continuously tune the dot radius from ≈40 nm to ≈10 nm. The concurrent observation of a large g-factor and strong intrinsic spin-orbit coupling two of the three ingredients for topological superconductivity together with the electrostatically tunable quantum confinement and the simplicity of solution-based synthesis, positions Te NWs as a compelling candidate for gate-defined spin qubits, chirality-resolved spintronics, and hybrid superconductor-semiconductor architectures targeting Majorana zero modes in an elemental van der Waals system.

## METHODS

### *Device fabrication.*

Devices were fabricated on heavily doped p-type silicon substrates capped with 285 nm of thermally grown $SiO_2$, which serves as the back-gate dielectric. The as-synthesized Te NWs were dispersed in methanol and subjected to a brief, gentle ultrasonication to separate individual wires before drop-casting onto the pre-cleaned substrate. After identifying suitable NWs by optical microscopy, source and drain contacts were defined using maskless photolithography. The contact metallization consisted of Pd/Au (20/30 nm) deposited by electron-beam evaporation for the source and drain electrodes, while Cr/Au (5/75 nm) was used for the bond pads to ensure robust wire-bonding adhesion. For the dual-gated devices (Figure 6), a 40-nm-thick hexagonal boron nitride (hBN) flake was transferred onto the NW channel as the top-gate dielectric, followed by patterning and metallization of the top-gate electrode.

***Transmission Electron Microscope imaging.***

Microstructural characterization was carried out using a Thermo Scientific Spectra 300 S/TEM (Thermo Fisher Scientific) operated at 200 kV. The instrument is an aberration-corrected field-emission gun transmission electron microscope. Low-magnification TEM imaging was used to examine the morphology and dimensions of the tellurium nanowires, while atomic-resolution HAADF-STEM imaging was performed to investigate their crystal structure and growth orientation.

***Electrical and magneto-transport measurements.***

Electrical transport measurements were performed by applying a finite DC bias across the source and drain electrodes using digital-to-analog converters (DACs). The resulting current was amplified by a current preamplifier, and the differential conductance dI/dV was obtained by numerical differentiation of the measured *I–V* characteristics. Low-temperature measurements were carried out in a dilution refrigerator with a base temperature of 10 mK, equipped with a superconducting solenoid magnet for magneto-transport studies. The magnetic field was applied either in-plane (at ~ 43° with respect to the NW c-axis) or out-of-plane (perpendicular to the substrate) to extract the anisotropic g-factors and spin-orbit gap reported in this work.

**ASSOCIATED CONTENT**

**Supporting Information.** Additional morphological and dimensional characterization of hydrothermally grown Te NWs by SEM and AFM (Figure S1); reproducibility of Fabry–Pérot interference signatures across multiple low-resistance devices with $R_{RT}$ < 30 kΩ (Figure S2); reproducibility of Coulomb-blockade signatures across multiple high-resistance devices with $R_{RT}$ ≥ 30 kΩ (Figure S3) (PDF).

## AUTHOR INFORMATION

### Corresponding Author

*Dharmraj Kotekar-Patil – Department of Physics, University of Arkansas, Fayetteville, Arkansas 72701, United States; MonArk NSF Quantum Foundry, University of Arkansas; Email: dk030@uark.edu

### Authors

Suresh Ghimire – Department of Physics, University of Arkansas, Fayetteville, Arkansas 72701, United States

Mohammad Hafijur Rahaman – Department of Physics, University of Arkansas, Fayetteville, Arkansas 72701, United States

Nathan Tanner Sawyers – Department of Physics, University of Arkansas, Fayetteville, Arkansas 72701, United States

Madan Bhandari – Department of Physics, University of Arkansas, Fayetteville, Arkansas 72701, United States

Pawan Kumar – Institute of Materials Research and Engineering, A*STAR, Singapore 138634, Republic of Singapore

Zainul Aabdin – Institute of Materials Research and Engineering, A*STAR, Singapore 138634, Republic of Singapore

Hugh O. H. Churchill – Department of Physics, University of Arkansas, Fayetteville, Arkansas 72701, United States; MonArk NSF Quantum Foundry, University of Arkansas

### Notes

The authors declare no competing financial interest.

## ACKNOWLEDGMENTS

This research is supported by MonArk NSF Quantum Foundry supported by the National Science Foundation Q-AMASE-i program under NSF Award No. DMR-1906383, and AFRL under agreement number FA8750-24-1-1019.

# SUPPORTING INFORMATION

## Fabry-Pérot Interference, g-factor Anisotropy, and Gate-Tunable QDs in Chiral Tellurium NWs

Suresh Ghimire[1], Mohammad Hafijur Rahaman[1], Nathan Tanner Sawyers[1], Madan Mohan Bhandari[1], Gokul Acharya[1], Syed Zulfiqar Hussain Shah[2,3], Iris Nandhakumar[3], Pawan Kumar[2], Zainul Aabdin[2], Hugh O. H. Churchill[1,4,5], and Dharmraj Kotekar-Patil[1,4,5*]

[1]*Department of Physics, University of Arkansas, Fayetteville, Arkansas 72701, United States*

[2]*Institute of Materials Research and Engineering, Agency for Science Technology and Research (A*STAR), Singapore 138634, Republic of Singapore*

[3] *Department of Chemistry, University of Southampton, Southampton SO17 1BJ, UK*

[4]*MonArk NSF Quantum Foundry, University of Arkansas, Fayetteville, Arkansas 72701, United States*

[5]*Arkansas Materials Institute, Fayetteville, Arkansas 72701, United States*

** Corresponding author: dk030@uark.edu*

### S1. Morphology and Dimensional Characterization

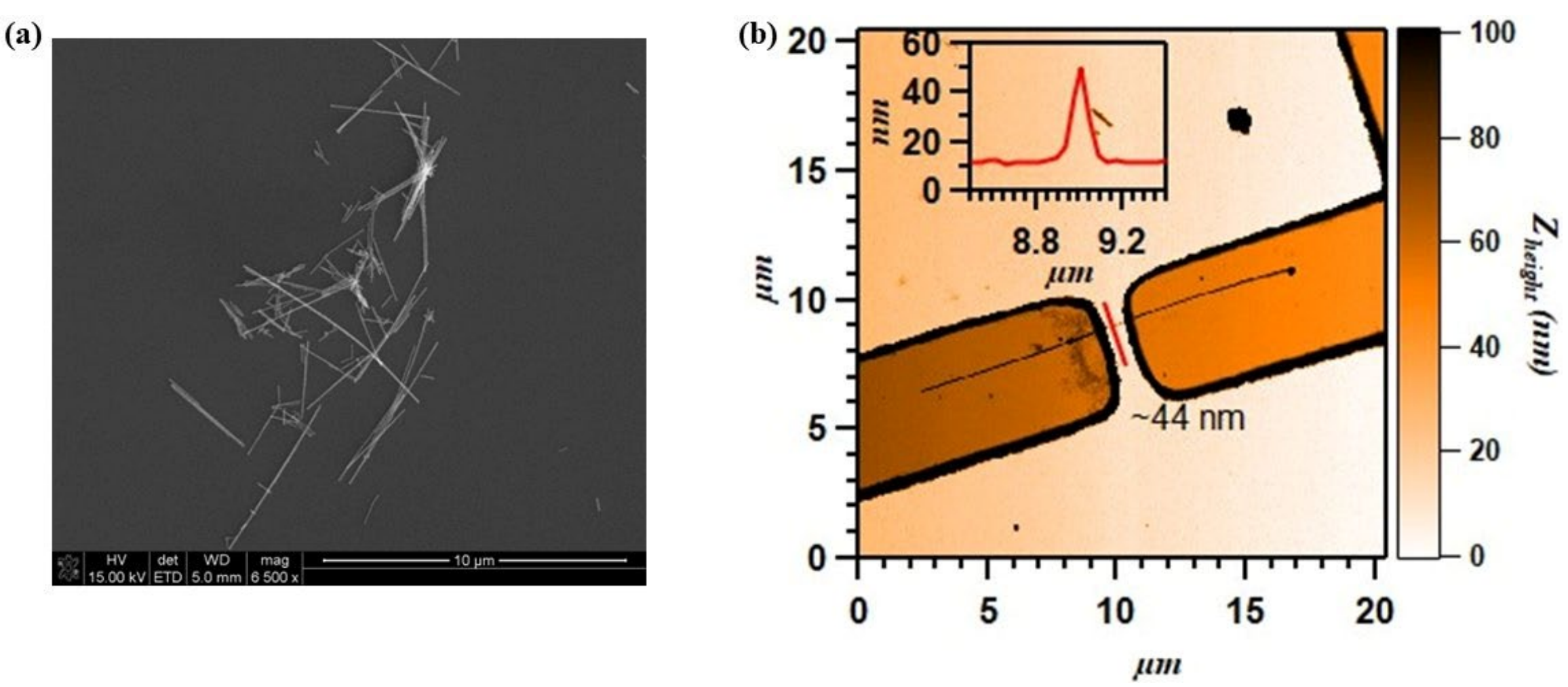

**Figure S1. SEM and AFM characterization of hydrothermally grown Te NWs.** (a) Wide-area SEM image showing as-grown Te NWs dispersed on $SiO_2$/Si substrate (scale bar 10 μm). The NWs exhibit typical lengths of tens of micrometers and diameters in the range of 20–100 nm. (b) AFM height map of a single Te NW (color scale: 0–100 nm). The NW shows uniform diameter along its length. Inset: cross-sectional height profile extracted along the dashed line, yielding a diameter of ≈44 nm.

## S2. Fabry–Pérot Interference Reproduced Across Multiple Devices

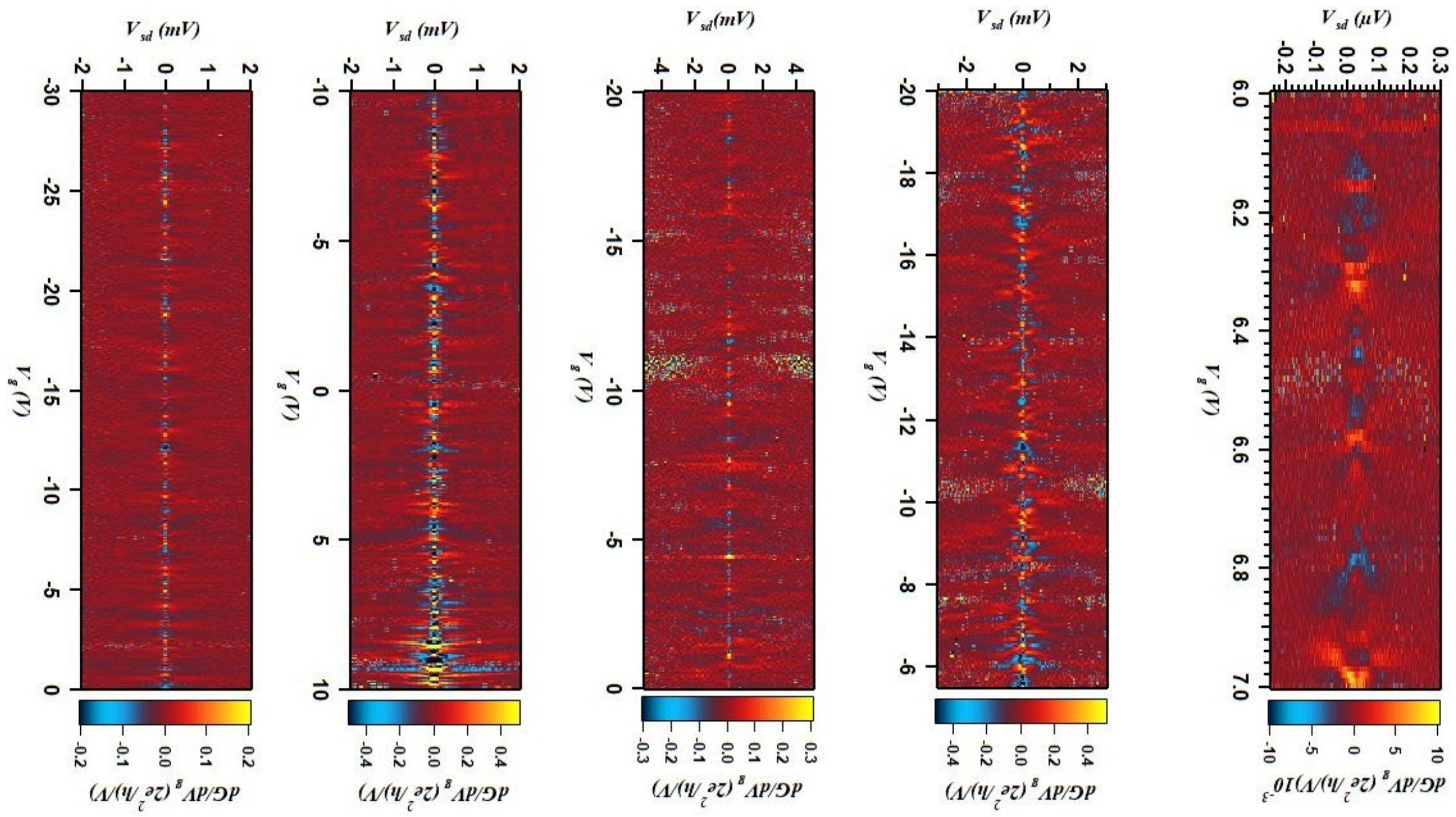

**Figure S2. Fabry–Pérot interference signatures observed in multiple Te-NW devices.** Five panels showing $dG/dV_g(V_{sd}, V_g)$ measured at base temperature for devices with $R_{RT}$ < 30 kΩ. Each panel corresponds to a different device and gate-voltage window, demonstrating the reproducibility of Fabry–Pérot interference across multiple low-resistance devices. The panels show varying degrees of checkerboard clarity and quasi-periodic conductance oscillations characteristic of phase-coherent resonant transmission through the NW channel. The consistent appearance of FP signatures across all low-resistance devices confirms that $R_{RT}$ < 30 kΩ is a robust threshold for accessing the quasi-ballistic transport regime.

## S3. Coulomb Blockade Reproduced Across Multiple Devices

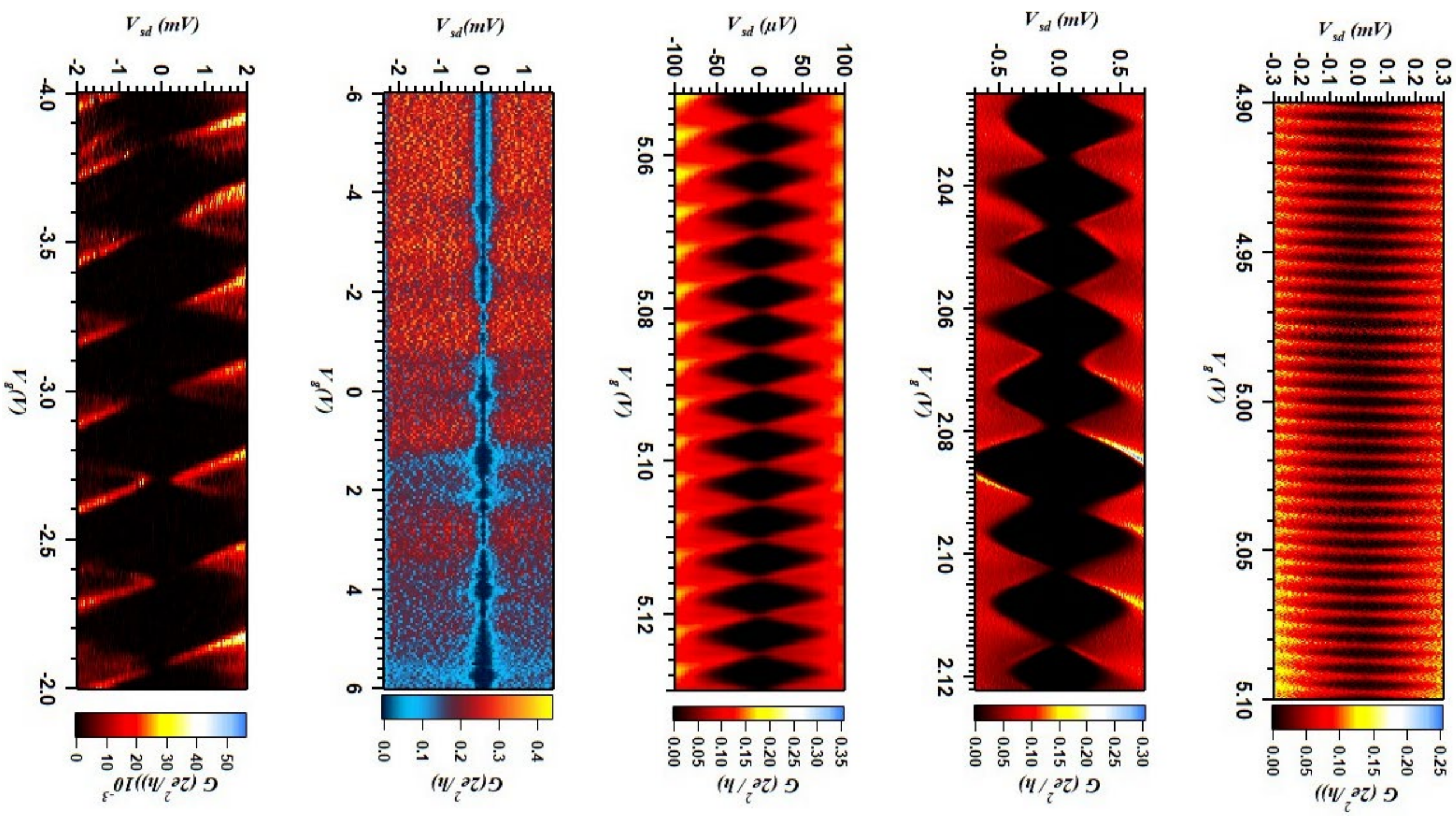

**Figure S3. Coulomb-blockade signatures observed in multiple Te-NW devices.** Five panels showing transport signatures at base temperature for devices with $R_{RT} \geq 30$ kΩ. Each panel corresponds to a different device and gate-voltage window, demonstrating the reproducibility of Coulomb-blockade behavior across multiple high-resistance devices. The consistent observation of diamond-shaped regions of suppressed conductance in all high-resistance devices confirms that $R_{RT} \geq 30$ kΩ displays a Coulomb blockade regime and that quantum dot formation is a reproducible feature of Te NW devices in this regime.